\documentclass[
    ,final            
  ]
  {aipproc}

\layoutstyle{6x9}

\begin{document}

\title{Late-Time X-ray Flares during GRB Afterglows: Extended Internal Engine Activity}


\author{A. D. Falcone}{
  address={Department of Astronomy \& Astrophysics, 525 Davey
  Lab., Penn. State University, University Park, PA 16802, USA},
  email={afalcone@astro.psu.edu}
}
\author{D. N. Burrows}{
  address={Department of Astronomy \& Astrophysics, 525 Davey
  Lab., Penn. State University, University Park, PA 16802, USA}
}
\author{P. Romano}{
  address={INAF-Osservatorio Astronomico di Brera, Via Bianchi 46, 23807 Merate, Italy}
}
\author{S. Kobayashi}{
  address={Department of Astronomy \& Astrophysics, 525 Davey
  Lab., Penn. State University, University Park, PA 16802, USA},
  altaddress={Astrophysics Research Institute, Liverpool John Moores University, Birkenhead CH41 1LD, UK}
}
\author{D. Lazzati}{
  address={JILA, University of Colorado, Boulder, CO 80309, USA}
  }
\author{B. Zhang}{
  address={Department of Physics, University of Nevada, Las Vegas, NV}
}
\author{S. Campana}{
  address={INAF -- Osservatorio Astronomico di Brera, Merate, Italy}
}
\author{G. Chincarini}{
  address={INAF -- Osservatorio Astronomico di Brera, Merate, Italy},
  altaddress={Universit\`a degli studi di Milano-Bicocca, Dipartimento di Fisica, Milano, Italy}
}
\author{G. Cusumano}{
  address={INAF- Istituto di Fisica Spazialee Fisica Cosmica sezione di Palermo, Palermo, Italy}
}
\author{N. Gehrels}{
  address={NASA Goddard Space Flight Center, Greenbelt, MD}
}
\author{P. Giommi}{
  address={ASI Science Data Center, via Galileo Galilei, 00044 Frascati, Italy}
}
\author{M. R. Goad}{
  address={Department of Physics and Astronomy, University of Leicester, Leicester, UK}
}
\author{O. Godet}{
  address={Department of Physics and Astronomy, University of Leicester, Leicester, UK}
}
\author{J. E. Hill}{
  address={USRA, 10211 Wincopin Circle, Suite  500, Columbia, MD, 21044-3432, USA},
  altaddress={NASA Goddard Space Flight Center, Greenbelt, MD}
}
\author{J. A. Kennea}{
  address={Department of Astronomy \& Astrophysics, 525 Davey
  Lab., Penn. State University, University Park, PA 16802, USA}
}
\author{P. M\'{e}sz\'{a}ros}{
  address={Department of Astronomy \& Astrophysics, 525 Davey
  Lab., Penn. State University, University Park, PA 16802, USA},
  altaddress={Department of Physics, Penn. State University, University Park, PA 16802, USA}
}
\author{D. Morris}{
  address={Department of Astronomy \& Astrophysics, 525 Davey
  Lab., Penn. State University, University Park, PA 16802, USA}
}
\author{J. A. Nousek}{
  address={Department of Astronomy \& Astrophysics, 525 Davey
  Lab., Penn. State University, University Park, PA 16802, USA}
}
\author{P. T. O'Brien}{
  address={Department of Physics and Astronomy, University of Leicester, Leicester, UK}
}
\author{J. P. Osborne}{
  address={Department of Physics and Astronomy, University of Leicester, Leicester, UK}
}
\author{C. Pagani}{
  address={Department of Astronomy \& Astrophysics, 525 Davey
  Lab., Penn. State University, University Park, PA 16802, USA}
}
\author{K. Page}{
  address={Department of Physics and Astronomy, University of Leicester, Leicester, UK}
}
\author{G. Tagliaferri}{
  address={INAF-Osservatorio Astronomico di Brera, Via Bianchi 46, 23807 Merate, Italy}
}
\author{the Swift XRT Team}{
  address={}
}

\begin{abstract}
Observations of gamma ray bursts (GRBs) with {\it{Swift}} produced the initially surprising result that many bursts have large X-ray flares superimposed on the underlying afterglow.  These flares were sometimes intense, rapid, and late relative to the nominal prompt phase.  The most intense of these flares was observed by XRT with a flux $>500\times$ the afterglow.  This burst then surprised observers by flaring again after $>10000$ s.  The intense flare can be most easily understood within the context of the standard fireball model, if the internal engine that powers the prompt GRB emission is still active at late times.  Recent observations indicate that X-ray flares are detected in $\sim$1/3 of XRT detected afterglows.  By studying the properties of the varieties of flares (such as rise/fall time, onset time, spectral variability, etc.) and relating them to overall burst properties, models of flare production and the GRB internal engine can be constrained.
\end{abstract}

\maketitle


\section{Introduction}

Since its launch on 2004 November 20, {\it{Swift}} \citep{geh04} has provided detailed measurements of numerous gamma ray bursts (GRBs) and their afterglows with unprecedented reaction times.  By detecting burst afterglows promptly, and with high sensitivity, the properties of the early afterglow and extended prompt emission can be studied in detail for the first time.  This also facilitates studies of the transition between the prompt emission and the afterglow.  The rapid response of the pointed X-ray Telescope (XRT) instrument \cite{bur05a} on {\it{Swift}} has led to the discovery that large X-ray flares are common in GRBs and occur at times well after the initial prompt emission. 

While there are still many unknown factors related to the mechanisms that produce GRB emission, the most commonly accepted model is that of a relativistically expanding fireball with associated internal and external shocks \citep{mes97}.  In this model, internal shocks produce the prompt GRB emission. Observationally, this emission typically has a timescale of $\sim30$ s for long bursts and $\sim$0.3 s for short bursts \citep{mee96}.  The expanding fireball then shocks the ambient material to produce a broadband afterglow that decays quickly (typically as ${\propto}t^{-\alpha}$).  When the Doppler boosting angle of this decelerating fireball exceeds the opening angle of the jet into which it is expanding, then a steepening of the light curve (jet break) is also predicted \citep{rho99}.  For a description of the theoretical models of GRB emission and associated observational properties, see \citet{mes02}, \citet{zha04}, \citet{piran05}, and \citet{van00}.  For an alternative explanation that describes both the prompt emission and the afterglow emission with a forward shock, see \citet{der99}.

With the advent of recent {\it{Swift}}-XRT observations of many large flares at various times after the burst, it is clear that a new constraint on GRB models is available to us.  We now know that the few previous observations of relatively small flux increases \cite{pir05,int03} did not provide a complete picture of the X-ray flaring activity during and following GRBs.  Recent observations by XRT indicate that flares are common, that they can have a fluence comparable to the initial prompt emission, and that they have various timescales, spectra, and relative flux increase factors \cite{bur05b, fal06, rom06, bur06}.  By studying the properties of these flares and by delving into the details of the GRB models, the nature of the X-ray flares, and possibly the GRB internal engine, may be elucidated.

\section{Overall XRT Observations}

As of 27 December 2005, {\it{Swift}}-BAT detected and imaged 95 GRBs, which extrapolates to a rate of $\sim100/year$.  {\it{Swift}} slewed to 80 of these bursts within 200 ks, and 74$\%$ of these observations resulted in detections of an X-ray afterglow.  XRT slewed promptly to 59 of these bursts within 350 s, and 95$\%$ of these observations resulted in detections of an X-ray afterglow.  From the sample of 56 bursts with prompt slews and detections, more than 24 of them have significant detections of X-ray flares at late times, relative to the nominal prompt emission time frame.  In short, $>25\%$ of all {\it{Swift}}-BAT detected bursts have significant X-ray flares, and $>43\%$ of the bursts with a prompt XRT detection have significant X-ray flares.

\section{A Few Remarkable Flaring GRBs}

The flaring GRBs discussed below are just a small subset of those observed so far.  A more comprehensive sample will be published soon in two forthcoming papers.

\subsection{XRF 050406}

XRF 050406 was the first {\it{Swift}} burst with flaring that was clearly significant, independent of any supporting observations \cite{rom06,bur05b}.  It is worth mentioning that GRB050219a exhibited flaring, but confidence was not achieved until the higher significance detections of flaring from XRF 050406, and then GRB 050502B.  XRF 050406 has a flare with a peak at about 210 s after the BAT trigger time.  The flare rises above the underlying power law decay by a factor of $\sim6$.  When the underlying power law decay, which has a temporal decay index of $1.58\pm0.17$, is subtracted from the flare data, the rise and fall of the flare are nearly symmetric with temporal power law indices of $\pm6.8$.  The ${\delta}t/t$ for this flare is $\sim0.2$.  The underlying decay curve before and after the flare are consistent with a single temporal power law decay.  This flare did not provide enough photons to perform a detailed spectral analysis, but from plotting the band ratio, it could be seen that the flare had a harder spectrum at the onset which softened back to that of the underlying afterglow as the flare decayed \cite{rom06}.  XRF 050406 is also notable since it was an X-ray Flash, rather than a classic GRB.  This common feature of XRFs and GRBs suggests a potential link between the two classes.

\subsection{GRB 050502B}

\begin{figure}
\centering
\includegraphics[width=0.5\linewidth]{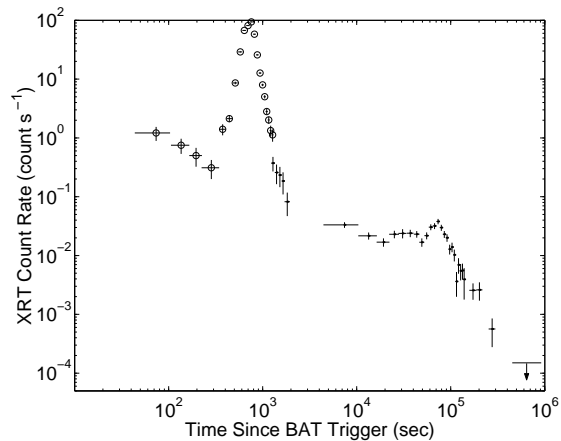}
\caption{X-ray light curve of GRB 050502B.  Open circles are window timing mode data, and dots are photon counting mode data.  For details, see \citet{fal06}.
\label{fig:050502B_lc}}
\end{figure}

GRB 050502B is a prime example of a GRB with at least one large flare at late times after the cessation of the initial prompt emission detected by BAT \cite{fal06, bur05b}.  The light curve from XRT data is shown in Figure~\ref{fig:050502B_lc}.  A giant flare, with a flux increase by a factor of $\sim$500, was observed using XRT.  The fluence during the giant flare, (1.2 $\pm$ 0.05) $\times 10^{-6}$ erg cm$^{-2}$ in the 0.2 -- 10 keV band, was slightly above that during the initial prompt emission detected by BAT.  The flare rises to a sharp peak at 743 $\pm$ 10 s, but this appears to be on top of a broader peak that extends from 640 $\pm$ 20 s to 790 $\pm$ 20 s.  In the hard band (1 -- 10 keV), there is significant time structure within the peak of the giant flare itself.  During the flare, the spectrum can be fit best by an absorbed cutoff power law (or Band function) \cite{ban93}, rather than a simple absorbed power law,  which fits the underlying afterglow nicely.  For details, see \citet{fal06}.  The spectral index hardens significantly during the flare (with a cutoff energy of $\sim$2.5 keV in the XRT band) before returning back to a softer and more typical afterglow spectrum after the flare has ended.  Before and after the flare, the temporal decay of the underlying afterglow can be fit well with a single power law $\sim t^{-0.8 \pm 0.2}$.  At much later times, between (1.9 $\pm$ 0.3) $\times10^{4}$ s and (1.1 $\pm$ 0.1) $\times10^{5}$ s, there are two broad bumps (or possibly one broad bump with some structure).  These bumps are notable in themselves since they could be more flaring, or they could be due to a combination of flaring and energy injection into the forward shock.

\subsection{GRB 050607}

GRB 050607 is notable due to the fast rise of one of its multiple flares \cite{pag06}.  The second, and brightest flare, had a peak at $\sim$310 s.  To borrow a term from prompt GRB descriptions, this flare was FRED-like (fast-rise, exponential-decay), with a very steep rise.  The temporal power law index was $\sim$22 if one placed $t_{0}$ at the burst trigger time, and it was $\sim$4.1 if one places $t_{0}$ at the time of the flare onset \cite{pag06}.  The ${\delta}t/t$ for this flare is $\sim0.2$.

\subsection{Flaring Short bursts}

The exceptional short burst, GRB 050724, exhibited significant flaring detected by XRT (for details, see \citet{bar05a}).  There were several flare-like features.  In particular, there is the broad bump detected with a peak at $\sim5\times10^{4}$ s.

It is also possible that GRB 051227, which has a significant X-ray flare peaking at $\sim110$ s, is a short burst \cite{bar05b}.  However, there is some ambiguity in its characterization as short or long.

\subsection{Flaring from High Redshift bursts}

\begin{figure}
\centering
\includegraphics[width=0.6\linewidth,clip,bb=95 389 457 630]{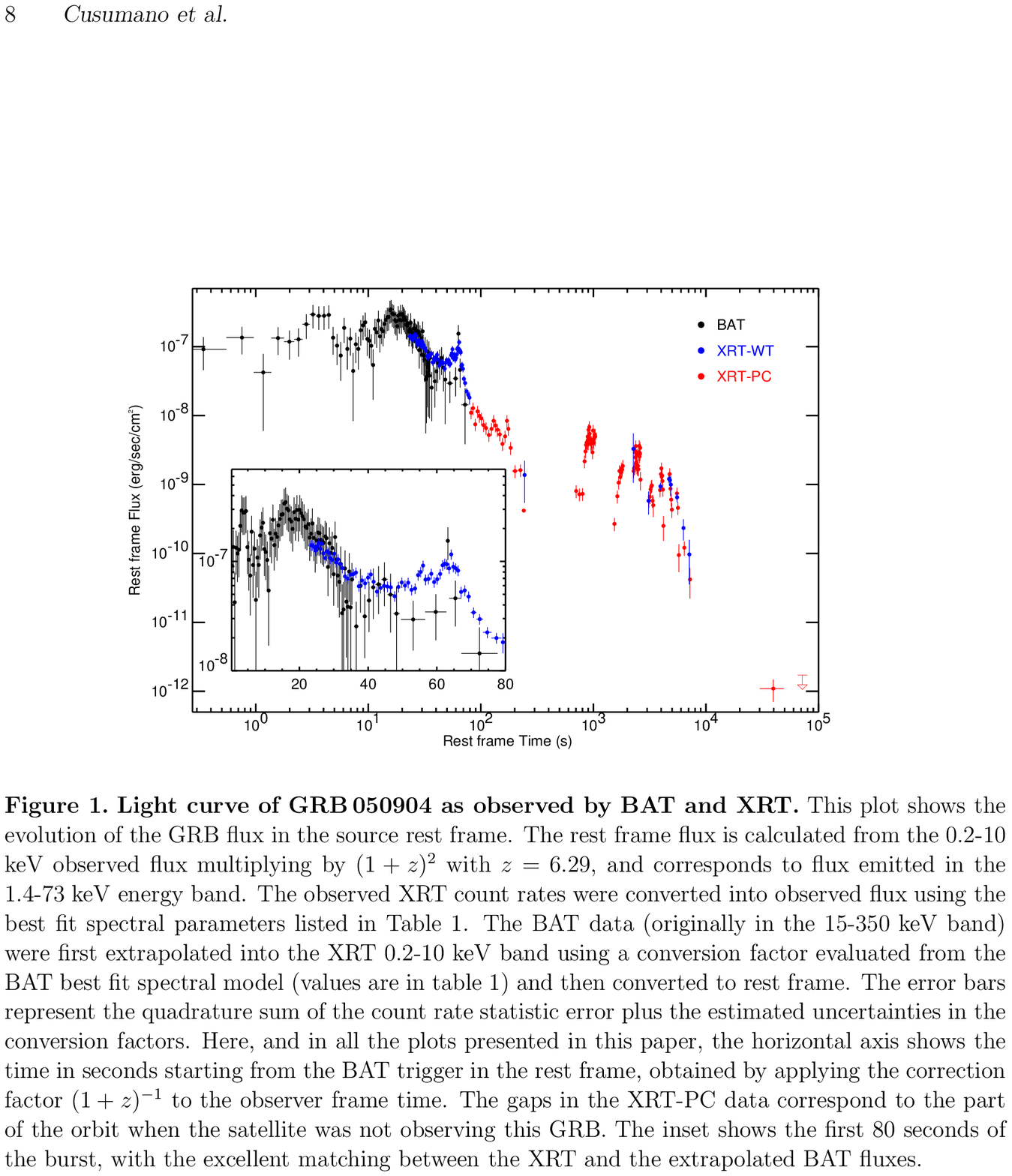}
\caption{Background-subtracted X-ray light curve of GRB~050904, transformed into the rest frame for z=6.29.  The BAT data points are from an extrapolation into the XRT energy range (black points, from 0-75~s) superimposed on the XRT light curve (20-10,000~s).  For details, see \citet{cus06}.
\label{fig:050904lc}}
\end{figure}

GRB 050904, at a redshift of 6.29, is the most distant GRB detected to date.  This burst has a very interesting X-ray light curve (see Figure~\ref{fig:050904lc}) with many flares superimposed on top of the underlying temporal decay, and on top of one another \cite{cus06}.  Even after a transformation of the light curve into the rest frame of the GRB, there is significant flaring at times as late as $\sim$5000 s.

In addition to GRB 050904, there are other moderately high redshift bursts with multiple flares (e.g. GRB 050730 at $z\sim4$).  As more data arrive from flaring GRBs, it will be interesting to compare all of the redshift corrected rest frame light curves.  While the high redshift certainly makes the emission more extended in the observer frame, these bursts are remarkably sporadic at late times, even in the rest frame.

\section{Discussion and Conclusions}

It is clear that we now have a recently realized characteristic of GRBs that can be used to probe their nature.  The X-ray flares have myriad characteristics.  Many have very fast rises and decays, whereas others are relatively gradual.  Some occur at early times, along with the nominal prompt emission detected by BAT, whereas others occur at very late times ($\sim10^{5}$ s).  They occur during all of the underlying decay curve phases (see \citet{nou06, zha06} for discussion of decay phases), with the possible exception of post-jet-break times.  Some of the flares are huge, whereas others are small bumps.  Some GRBs exhibit many flares, whereas other GRBs have only one.

A large fraction of the flares have several characteristics that point towards continued internal engine activity.  These characteristics include: 1) The temporal decay index before and after many (but not all) flares are identical, indicating that the afterglow had already begun before the flare, 2) the rise time and decay time of the flares are frequently very fast (${\delta}t/t \ll 1$), thus the flare is difficult (although not impossible) to explain with mechanisms associated with the external shock (see \citet{iok05,zha06} for discussion) , 3) there is even faster time structure near the peaks of some of the flares, 4) the spectra during some flares are represented better by a Band function or cutoff power law model, rather than a simple power law, similar to the nominal prompt emission, 5) the hardness before and after some flares is consistent with an afterglow that has already begun before the flare and continues with approximately the same spectral index after the flare, whereas the spectra during some flares are frequently harder than the underlying afterglow.  A final piece of supporting evidence for the restarting of the central engine is that the decay parameters following flares (and BAT prompt emission) usually imply a $t_{0}$ that is consistent with the onset of the event, when the decay is interpreted as being dominated by the curvature effect; for details, see \citet{lia06}.

For at least some GRBs with flares, continued internal engine activity is likely, but some flares allow for the possibility of external shock processes, within the framework of the standard fireball model.  It is beyond the scope of this paper to address particular extended internal engine models that can explain these flare observations.  However, it is important to note that any such models must be capable of emission at very late times ($>10^{4}$ s), sporadic and repeated emission to explain multiple flares, very fast rise/decay times, and total energy input comparable to that of the initial prompt emission.

Studies of the overall properties of a sample of many flaring GRBs are necessary to truly characterize their nature, and to determine if there are classes.  Results from these studies are forthcoming.


\begin{theacknowledgments}

This work is supported at Penn State by NASA contract NAS5-00136; at the Univ. of Leicester by the Particle Physics and Astronomy Research Council under grant PPA/Z/S/2003/00507; and at OAB by funding from ASI under grant I/R/039/04.

\end{theacknowledgments}

\bibliographystyle{aipproc}   

\end{document}